\documentstyle[12pt]{article}
\newcommand{\beq}{\begin{equation}}
\newcommand{\eeq}{\end{equation}}
\newlength{\bredde}
\def\slash#1{\settowidth{\bredde}{$#1$}\ifmmode\,\raisebox{.15ex}{/}
\hspace*{-\bredde} #1\else$\,\raisebox{.15ex}{/}\hspace*{-\bredde} #1$\fi}
\begin{document}
\vspace{1cm}

\title{\Large{Smooth Bosonization II: The Massive Case}}
\vspace*{1cm}

\author{{\sc P.H. Damgaard}             \\
CERN -- Geneva \\
{}~~ \\
{\sc H.B. Nielsen} and {\sc R. Sollacher} \\
The Niels Bohr Institute \\
Blegdamsvej 17,
DK-2100 Copenhagen, Denmark}

\maketitle
\vfill
\begin{abstract}
The (1+1)-dimensional bosonization relations for fermionic mass terms
are derived by choosing a specific gauge in an enlarged gauge-invariant theory
containing both fermionic and bosonic fields. The fermionic part of the
generating functional subject to the gauge constraint can be cast into the form
of a strongly coupled Schwinger model, which can be solved exactly. The
resulting bosonic
theory coupled to the scalar sources then exhibits directly the bosonic
counterparts of the fermionic densities $\bar{\psi}\psi$ and
$\bar{\psi}\gamma_5\psi$.
\end{abstract}

\vspace{80mm}

\begin{flushleft}
CERN--TH-6563/92 \\
July 1992 \\
\end{flushleft}



\vfill
\newpage
In a previous paper \cite{us}, we have shown
how the Abelian bosonization relations for the vector and axial vector
currents in two dimensions
\cite{Susskind,Col:75,Mandst:75} can be derived from a novel perspective,
based on a new local ``bosonization gauge symmetry" \cite{talk}. The idea
is, briefly stated, that bosonic and fermionic formulations of the same
theory should be understood as two different gauge fixings of a larger
gauge-invariant action containing both bosonic and fermionic fields.
Equivalence between certain purely bosonic and purely fermionic formulations
should then amount to the usual gauge-fixing independence of S-matrix
elements in gauge theories. Although this idea is appealing, it is a
non-trivial matter to show in detail that it can be carried through. One
needs two ingredients: (a) The local bosonization gauge symmetry and
the ``larger" gauge-invariant action, and, (b) a smooth interpolating
gauge-fixing function which can bring one continuously from a ``boson
gauge" to a ``fermion gauge" by tuning the gauge-fixing parameter.

In the usual path-integral approach to bosonization \cite{many,Russian,Rudi},
it is clear that chiral fermion determinants play a fundamental r\^{o}le.
The idea of \cite{us} was therefore to start with a fermionic theory, and then
promote chiral rotations to a local chiral gauge symmetry. Using the
general scheme of \cite{AlfDam:90}, this can be done without changing the
physical content of the theory through the introduction of a collective
field $\theta(x)$. This collective field -- in essence the chiral phase
of the fermions --  turns out to be the bosonized field in a suitable
gauge. In \cite{us} we showed how to
introduce a particular gauge-fixing function
$\Phi(x)$ depending on a gauge parameter $\Delta$ in such a way that one
interpolates smoothly between purely fermionic formulations ($\Delta = 0$)
and purely bosonic formulations ($\Delta = 1$). By comparing the couplings
to external sources we recovered the usual bosonization relations for the
currents.
One of the most interesting aspects of this gauge-invariant approach
to (1+1)-dimensional bosonization is that it demonstrates that these
known equivalences are but two extreme cases of a continuum of
equivalent theories that contain, in general, both fermions and bosons
interactively. For this reason we dubbed our scheme {\em smooth bosonization}.

Although the current bosonization relations $\bar{\psi}\gamma_{\mu}\psi
\sim - \pi^{-1/2} \epsilon_{\mu\nu}\partial^{\nu} \theta$ and
$\bar{\psi}\gamma_{\mu}\gamma_5\psi \sim \pi^{-1/2}\partial_{\mu}\theta$
can be established in this manner by a fairly direct route, the
derivation of the much more subtle mass-term bosonization relations
\beq
\bar{\psi}\psi \sim M \cos(2\sqrt{\pi}\theta)~~ {\mbox{\rm and}}~~~~~
\bar{\psi}\gamma_5\psi \sim M \sin(2\sqrt{\pi}\theta)
\eeq
was only
outlined in \cite{us}. The purpose of this letter is to fill in this gap.

Since we have already shown in \cite{us} how to treat external vector and axial
vector current sources, we shall only consider the generating functional
\begin{eqnarray}
{\cal{Z}}[M_{\pm}] &~=~& \int {\cal{D}}[\bar{\psi},\psi] e^{i\int d^2x
{\cal{L}}(x)} \nonumber \\
{\cal{L}}(x) &~=~& \bar{\psi}(x)(i\slash{\partial} + M_+(x)P_+
+ M_-(x)P_-)\psi(x) ,
\label{eq:Zmass}
\end{eqnarray}
which for constant sources $M_{\pm}$ just corresponds to having scalar and
pseudoscalar mass terms. Here $P_{\pm} \equiv  (1\pm \gamma_5)$ are
the usual chiral projectors.

Although global chiral symmetry is broken explicitly by the sources, we
still have what at first sight looks like a trivial {\em discrete} chiral
symmetry,
\begin{eqnarray}
\psi(x) &~\to~& e^{in\pi\gamma_5}\psi(x) ~=~ (-1)^n \psi(x) \nonumber \\
\bar{\psi}(x) &~\to~& \bar{\psi}(x)e^{in\pi\gamma_5} ~=~ (-1)^n \bar{\psi}(x) ,
\label{eq:dis}
\end{eqnarray}
where $n$ is an integer. When we introduce a collective field $\theta(x)$
via a chiral rotation
\beq
\psi(x) ~=~ e^{i\theta(x)\gamma_5}\chi(x) ,
\eeq
this degree of freedom is therefore only defined globally modulo $n\pi$,
$i.e$, the gauge fixing $\delta$-function must be periodic. If, in the
notation of \cite{us}, we take the case $\Delta = 0$, this gauge-fixing
function
is $\Phi(x) = \theta(x)/\pi$, and hence
\beq
\delta(\Phi(x)) ~=~ \delta(\Phi(x) - n).
\label{eq:period}
\eeq
We can give a convenient representation of such a globally periodic
functional $\delta$-function by a Fourier transform:
\beq
\delta(\Phi(x)) ~=~ \sum_{k=-\infty}^{\infty}\int {\cal{D}}[b]_k
\exp\left[\int d^2x b(x) \Phi(x)\right] ,
\label{eq:delta}
\eeq
where the functional integral is performed over all $b$'s satisfying
the constraint
\beq
\frac{1}{\pi}\int d^2x b(x) = k ,
\label{eq:bindex}
\eeq
where $k$ is an integer.

The interpolating gauge-fixing function reads in detail, for general
$\Delta$ \cite{us}:
\beq
\Phi(x) ~=~ \Delta\int_{-\infty}^x d\xi^{\nu} \bar{\chi}(\xi)
\gamma_{\nu}\gamma_5\chi(\xi) + \frac{1-\Delta}{\pi}\theta(x) ,
\eeq
where, for convenience, we have shifted the arbitrary lower limit of
the line integral all the way to $-\infty$.\footnote{In a manner
somewhat reminiscent of the prescription used by Mandelstam
\cite{Mandst:75} in relating boson and fermion operators. Here, the
point $-\infty$ should simply be viewed as any point ``outside" the
space-time region we are considering. A more rigorous treatment could
either keep $x_0$, the lower limit of the line integral finite,
and then modify the analysis by the inclusion of zero modes, -- or
choose appropriate boundary conditions in a finite volume.}
The periodicity of the $\delta$-function constraint is not immediately
obvious if we choose $\Delta \neq 0$, since under the discrete chiral
transformation (\ref{eq:dis}), the contribution from the
$\theta$-term above now becomes $-(1-\Delta)n$. However, the line
integral {\em also} transforms under the discrete chiral rotation,
\beq
\Delta\int_{-\infty}^x d\xi^{\nu}\bar{\chi}\gamma_{\nu}\chi
 ~\to~ \Delta\int_{-\infty}^x d\xi^{\nu}\bar{\chi}\gamma_{\nu}\chi
- n\Delta,
\eeq
as can be seen by choosing an extremely slowly varying function, $\alpha(x)$,
tending to the limit of $n\pi$ everywhere (except at $-\infty$), and then
performing the corresponding regularized discrete chiral rotation
\beq
\bar{\chi} \to \bar{\chi}e^{i\alpha(x)\gamma_5}~,~~~~ \chi \to
e^{i\alpha(x)\gamma_5}\chi.
\eeq
Under this transformation, the line integral transforms on account of the
axial anomaly:
\beq
\Delta\int_{-\infty}^x d\xi^{\nu}\bar{\chi}\gamma_{\nu}\chi
 ~\to~ \Delta\int_{-\infty}^x d\xi^{\nu}\bar{\chi}\gamma_{\nu}\chi
- \frac{\Delta}{\pi}\alpha(x) ~\to~
\Delta\int_{-\infty}^x d\xi^{\nu}\bar{\chi}\gamma_{\nu}\chi- n\Delta.
\eeq
The two pieces therefore add up, leaving us again with the result
(\ref{eq:period}).

Using
the Fourier representation (\ref{eq:delta}), we can rewrite the exponent as an
interaction term coupling the fermions and the boson $\theta$ to a
vector potential $B_{\mu}(x)$ defined by
\beq
B_{\mu}(x) ~=~ \epsilon_{\mu\nu}\int d^2y b(y) \int_{-\infty}^x
d\xi^{\nu} \delta^{(2)}(\xi-x).
\eeq
The Pontryagin index density of this vector potential is
\beq
\frac{1}{\pi} \epsilon^{\mu\nu}\partial_{\mu}B_{\nu}(x)
{}~=~ - \frac{1}{\pi} b(x) ;
\eeq
by the requirement (\ref{eq:bindex}), this implies
\beq
- \frac{1}{\pi}\int d^2x \epsilon^{\mu\nu}\partial_{\mu}B_{\nu}(x)
{}~=~ k
\eeq
where the integer $k$ is the instanton number of the gauge potential $B_{\mu}$.
In general, this gauge potential can be decomposed as (see, $e.g.$, ref.
\cite{Jaye})
\beq
B_{\mu}(x) ~=~ k C_{\mu}(x) + \epsilon_{\mu\nu}\partial^{\nu}
\tilde{b}(x) + \partial_{\mu}\varphi(x),
\label{eq:decomp}
\eeq
but we can always remove the third term by a local phase rotation of the
fermion fields. This simply corresponds to choosing a gauge for $B_{\mu}(x)$.
The field $C_{\mu}(x)$ is a background which is only constrained to have
the volume integral of $\pi^{-1}\epsilon_{\mu\nu}\partial^{\nu}C^{\mu}(x)$
fixed, $i.e.$ it
carries topological number 1. We can choose $C_{\mu}$ such that
$\partial_{\mu}C_{\nu}-\partial_{\nu}C_{\mu}$ is constant \cite{Wipf}. The
relation of the measure defined by (\ref{eq:delta}) to those in terms of
$k,\tilde{b}$, respectively $B_\mu$ is given by
\beq
\sum_{k=-\infty}^{+\infty} \int {\cal D}[b]_k = \sum_{k=-\infty}^{+\infty} \int
{\cal D}'[\tilde{b}] |{\det}'(\partial^2)| = \sum_{k=-\infty}^{+\infty} \int
{\cal D}'[\tilde{b}] {\cal D}'[\varphi ] \delta (\varphi )
|{\det}'(\partial^2)| ,
\eeq
where the prime denotes that the zero-mode sector (with respect to
$\partial^2$)
is excluded. The last expression is nothing but the measure for the vector
field $B_\mu$ in the decomposition (\ref{eq:decomp}) for the gauge $\varphi =
0$.

Let us now add a term
\beq
\frac{1}{2g^2} b(x)^2 ~=~ - \frac{1}{4g^2} B_{\mu\nu}(x)B^{\mu\nu}(x)
\label{eq:kin}
\eeq
to the action. The field strength is
$B_{\mu\nu}(x) = \partial_{\mu}B_{\nu} - \partial_{\nu}B_{\mu}$.
Surprisingly, the dimensionful coupling constant
$g$ will turn out to play the r\^{o}le of an ultraviolet cut-off.
The effect of adding such a term is to smear the $\delta$-function
of the gauge-fixing into a Gaussian. The advantage is that, before
integrating over $\theta$, our
action now contains one part which is the Schwinger model coupled to an
external source. We can then directly make use of results
which have been established for that model \cite{Jaye}. Later, we of course
send the cut-off $g$ to infinity, thereby reducing the gauge-fixing function
to the original $\delta$-function constraint.

\section*{The case $\Delta = 1$: Bosonization}

Putting everything together, we can now represent our gauge-fixed Lagrangian as
\begin{eqnarray}
{\cal L} &=& \bar{\chi} \biggl( i\partial\hskip-5pt / - \partial\hskip-5pt /
\theta \gamma_5 + B\hskip-5pt / + M_+ e^{2i\theta} P_+ + M_- e^{-2i\theta} P_-
\biggr) \chi \cr
& & - \frac{1}{4g^2} B_{\mu\nu}B^{\mu\nu} + \frac{1}{2\pi} \partial_\mu \theta
\partial^\mu\theta + M_+(e^{2i\theta} -1) \frac{\kappa_1 (\Lambda )}{4\pi}\cr
& &  + M_-
(e^{-2i\theta} -1) \frac{\kappa_1 (\Lambda )}{4\pi} -\frac{1}{8\pi}
\biggl(M_+^2
(e^{4i\theta} -1) + M_-^2 (e^{-4i\theta} -1) \biggr) .
\label{eq:Lgf}
\end{eqnarray}
The additional terms depending on $M_\pm$ result from the introduction of the
field $\theta$ via a chiral transformation. The corresponding Jacobian of such
a transformation, calculated with a Pauli-Villars regularization, contains
also mass-dependent terms (see \cite{Ball:89,us}). With the
constants $c_i, k_i$ obeying the relations
\beq
c_1 k_1 + c_2 k_2 = 0 \quad , \quad c_1 + c_2 = 1  ,
\eeq
the cut-off--dependent constant $\kappa_1 (\Lambda )$ is defined as
\beq
\kappa_1 (\Lambda ) = \Lambda \sum_i c_i k_i \log k_i^2 .
\eeq
For convenience, and in order to compare our results
with the work of Dorn \cite{Dorn}, we
use a scheme where $c_1 = c_2 = \frac{1}{2}, k_1 = -k_2 =1$. With this choice
$\kappa_1 (\Lambda) =0$.

It should be noted that the terms $ M_\pm^2 (e^{\pm4i\theta} - 1)$
in eq. (\ref{eq:Lgf}) arise as a consequence of this particular regularization
scheme. As we shall see later, they correspond to contact terms of the
$\delta$-function kind in certain Green functions. Such $\delta$-function
contributions are not na\"{i}vely expected to occur in a regularized theory.
However, Pauli-Villars regularization is defined here as a subtraction scheme
of
{\em connected} $n$-point functions (see \cite{Ball:89}), not as a
regularization of
the free propagator itself. This leads to $\delta$-function contributions when
calculating quantities involving more than one loop, like the
two-loop contribution $M_{\pm}^2(e^{\pm4i\theta}-1)$ to the Jacobian mentioned
above. It will turn out that these contact terms
proportional to $M_\pm^2$ appear in the bosonized version of the generating
functional as well. One could try to avoid the occurrence of these terms by
using a different regularization method; however, some very
pleasant features would then have to be given up. In particular,
this Pauli-Villars regularization guarantees
that independent (commuting) functional derivatives of
${\cal Z}[V_\mu ,A_\mu ,M_\pm]$
yield the same result, irrespective of the order in which they are taken, as
was shown in ref. \cite{Ball:89}.

The Lagrangian (\ref{eq:Lgf}) looks almost like a Schwinger model coupled to
sources  $M_{\pm}(x)e^{\pm2i\theta(x)}$, except for the derivative coupling of
$\theta$  to the
fermions\footnote{And of course the additional part of the Lagrangian
depending only on $\theta$. Since $\theta$ is a dynamical field, the full
theory is not completely decoupled into separate sectors. But as we shall
henceforth perform a perturbative expansion in the terms
$\bar{\chi}M_{\pm}(x)\exp[\pm 2i\theta(x)]$, this has no consequences
for the following analysis.}. This coupling can be removed by the following set
of transformations:
\begin{eqnarray}
\chi &\to& e^{ia\theta\gamma_5} \chi  \cr
B_\mu &\to& B_\mu + (1+a) \epsilon_{\mu\nu} \partial^\nu \theta \cr
\theta &\to& (1+a)^{-1} \theta .
\end{eqnarray}
These transformations have to be carried out in the order shown. Here, $a$ is a
non-local operator:
\beq
a = - \partial^2 \left(\partial^2 + \frac{g^2}{\pi} \right)^{-1} .
\label{eq:defa}
\eeq
As a consequence the kinetic term for $\theta$ acquires an additional term,
which leads to a Pauli-Villars regularized propagator with regulator mass
$\frac{g}{\sqrt{\pi}}$. The Lagrangian reads as follows:
\begin{eqnarray}
{\cal L} &=& {\cal L}_{Sch} + {\cal L}_{int} + {\cal L}_{\theta} \cr
{\cal L}_{Sch} &=& \bar{\chi} (i\partial\hskip-6pt / + B\hskip-6pt / ) \chi -
\frac{1}{4g^2} B_{\mu\nu} B^{\mu\nu} \cr
{\cal L}_{int} &=& \bar{\chi} (M_+ e^{2i\theta} P_+ + M_- e^{-2i\theta}
P_-)\chi \cr
{\cal L}_{\theta} &=& + \frac{1}{2g^2} \partial_\mu \theta \left(\partial^2 +
\frac{g^2}{\pi}\right)\partial^\mu \theta  -\frac{1}{8\pi} \biggl(M_+^2
(e^{4i\theta} -1) + M_-^2 (e^{-4i\theta} -1) \biggr) .
\end{eqnarray}

Now we are in a position to derive an effective bosonic theory in terms of
the field $\theta$. Let us therefore treat ${\cal L}_{int}$ as a perturbation
of the Schwinger model part, $i.e.$
\beq
\langle \exp \biggl( i\int d^2x {\cal L}_{int} (x)\biggr) \rangle = \exp
\biggl( i\int d^2x \langle {\cal L}_{int} (x) \rangle + \frac{i^2}{2} \int d^2
x \int d^2y \langle {\cal L}_{int} (x) {\cal L}_{int} (y) \rangle_c + \ldots
\biggr)
\label{eq:cum}
\eeq
The expectation values have to be taken with respect to $\chi , \bar{\chi}$ and
$B_\mu$. The expression $\langle \ldots \rangle_c$ means the connected part of
the correlation function. The resulting effective bosonic action reads
\begin{eqnarray}
{\cal L}_{eff} &=& {\cal L}_{\theta} + {\cal L}_{M_\pm} \cr
{\cal L}_{M_\pm} &=& M_+ (x) e^{2i\theta (x) } \langle \bar{\chi} (x) P_+ \chi
(x) \rangle + (+\leftrightarrow -) \cr
& & +~ \frac{i}{2} \int d^2y M_+ (x) e^{2i\theta (x)} \langle \bar{\chi} (x)
P_+
\chi (x) \bar{\chi} (y) P_+ \chi (y) \rangle_c M_+ (y) e^{2i\theta (y)} \cr
& & +~ \frac{i}{2} \int d^2y M_+ (x) e^{2i\theta (x)} \langle \bar{\chi} (x)
P_+
\chi (x) \bar{\chi} (y) P_- \chi (y) \rangle_c M_- (y) e^{-2i\theta (y)} \cr
& & +~ (+\leftrightarrow -) \cr
& & +~ \ldots
\label{eq:Leff}
\end{eqnarray}
All constants independent of $\theta$ or $M_\pm$ have been dropped. They are
unimportant for our discussion because they only contribute to the
normalization of the functional integral.

We can now use known results for the Schwinger model, in particular the
cluster decomposition property \cite{Niels,Jaye}. First of all, for
the chiral condensates we obtain the well-known expression
\beq
\langle \bar{\chi} P_\pm \chi \rangle = \frac{g e^\gamma}{\sqrt{\pi} 4\pi} ,
\eeq
where $\gamma$ is Euler's constant. For the 2-point functions of fermion
bilinears we get (in
Euclidean space, and including the contributions from the Pauli-Villars
regulators for the fermions \cite{Jaye,Dorn}):
\beq
\langle  \bar{\chi} (x) P_\pm \chi (x)  \bar{\chi} (y) P_\pm \chi (y)\rangle_c
=
\frac{1}{4\pi} \delta (x-y) + \biggl( \frac{g e^\gamma}{\sqrt{\pi}4\pi}
\biggr)^2 \biggl( e^{-2K_0 (\frac{g}{\sqrt{\pi}} |x-y|)} -1 \biggr) ,
\eeq
and
\beq
\langle  \bar{\chi} (x) P_\pm \chi (x)  \bar{\chi} (y) P_\mp \chi (y) \rangle_c
 = \biggl( \frac{g e^\gamma}{\sqrt{\pi}4\pi} \biggr)^2 \biggl( e^{2K_0
(\frac{g}{\sqrt{\pi}} |x-y|)} \Theta(e^{\gamma}\Lambda|x-y| -1) -1 \biggr),
\eeq
valid up to terms of ${\cal{O}}(\Lambda^{-1})$. Here $K_0$ is a modified
Bessel function of the second kind.
Of course we take the limit
where $\Lambda$ is sent to infinity, and as a consequence the step-function
$\Theta(e^{\gamma}\Lambda|x-y| -1)$ above simply equals unity for
physical distances\footnote{That is, distances larger than $1/\Lambda$.}. The
scheme-dependent
$\delta(x-y)$ contribution from $\langle  \bar{\chi} (x) P_\pm \chi (x)
\bar{\chi} (y) P_\pm \chi (y)\rangle_c$ to the effective action of eq.
(\ref{eq:Leff}) just
cancels the $(1/8\pi)M_{\pm}(x)^2 e^{\pm4i\theta(x)}$-terms of ${\cal
L}_{\theta}$. This cancellation is already one highly non-trivial step towards
identifying the bosonization relations.

The cluster decomposition property (shown in \cite{Jaye} to hold even in this
{\em massless} Schwinger
model in which we are computing averages) states that
the connected part of the
correlation functions vanishes for distances $|x-y| \gg \frac{\sqrt{\pi}}{g}$.
This can be seen quite easily from the above expressions, taking into account
the properties of the Bessel function $K_0(z)$ at large argument.
If we take the limit $g \to \infty$ in order to recover the old
$\delta$-function gauge, all these connected correlation functions vanish
for all non-vanishing distances. It
may appear surprising that the gauge parameter $g$ plays such a crucial
r\^{o}le in simplifying the analysis; after all, any physical answer is
independent of this parameter. However, although we may recover the same
final results also for finite $g$, the corresponding equivalent bosonic
theory will in general be highly non-local, and not very illuminating.
As we shall see below, it is only in the limit $g \to \infty$ that we recover
the standard local bosonized action. So the most convenient bosonization
gauge remains the $\delta$-function choice of \cite{us}, which
corresponds precisely to $g \to \infty$.

A note should also be made here concerning the distance scales involved
at this point. We started out with an ultraviolet cut-off $\Lambda$
from the Pauli-Villars regularization of the original fermion theory.
Through the modified gauge-fixing function [the addition of a kinetic
energy term for the gauge potential (\ref{eq:kin})], we introduced what turned
out to be a Pauli-Villars regulator for the boson field $\theta(x)$. Finally,
we also need an infrared cut-off $\mu$ for the computation of certain
Green functions (see below). Although both ultraviolet cut-offs $\Lambda$
and $g$ are eventually taken to infinity, we are always performing the
analysis in a certain distance regime set by these two scales. First of
all, from the beginning we clearly restrict the whole analysis to distance
scales $|x-y| \gg 1/\Lambda$.
The cluster decomposition \cite{Jaye} sets in for distances $|x-y|> 1/g$.
Taking the limit $g\to \infty$
guarantees that all higher connected correlation functions vanish at all
physical distances. However, if we wish to retain a finite cut-off $g$, then
all terms in the expansion contribute to the effective bosonic theory,
suppressed only by powers of $g^{-1}$.

Finally, adding all the above ingredients and rescaling $\theta$ by
$\sqrt{\pi}$, we arrive at
a bosonized theory described by a Lagrangian
\beq
{\cal L}_{bos} = \frac{\pi}{2g^2} \partial_\mu \theta \left(\partial^2 +
\frac{g^2}{\pi}\right)\partial^\mu \theta  + M_+ \frac{g e^\gamma}{\sqrt{\pi}
4\pi}
e^{2i\sqrt{\pi}\theta} + M_- \frac{g e^\gamma}{\sqrt{\pi} 4\pi}
e^{-2i\sqrt{\pi}\theta} -
\frac{1}{8\pi} ( M_+^2 + M_-^2 ) ,
\label{eq:Lbos}
\eeq
where the limit $g\to\infty$ is to be taken.\footnote{It should again
be emphasized that there are corrections to this simple effective action,
down by powers of $g^{-1}$. The shown terms should really be viewed as
the result of the limit $g \to \infty$. Everything is here expressed in
terms of bare quantities.} This coincides with the
result of Dorn \cite{Dorn} if we identify $g/\sqrt{\pi}$ with his
Pauli-Villars regulator mass $M_S$. Indeed, the kinetic term for $\theta$ is
precisely the corresponding Pauli-Villars regularized kinetic energy.
Taking the sources $M_{\pm} = m$ to be constant, this is the Sine-Gordon
action with a Pauli-Villars mass $g/\sqrt{\pi}$ and
one particular normal-ordering prescription in the operator
formalism \cite{Col:75}. We hope that the present derivation of
the bosonization relations (1) for fermionic mass terms has
underlined some of the subtleties behind the statement that
the massive Thirring model is to be viewed as equivalent to the
2D Sine-Gordon theory. But the path-integral manipulations of the
present paper can also be viewed as a much simplified
derivation of the classic result of Coleman \cite{Col:75}.

\section*{Conclusion}

We have shown how to regain the usual mass-term bosonization identities
by means of the gauge $\Phi$ with $\Delta=1$. The case $\Delta=0$ of course
trivially yields the purely fermionic formulation. As in previous
work \cite{Col:75}, the equivalent bosonic Lagrangian has been derived
only through its perturbative expansion. However, in contrast with
\cite{Col:75}
and other similar approaches using path integral methods \cite{many},
we do not explicitly need an order-by-order comparison
in the perturbative expansion. Using the established cluster decomposition
property of the Schwinger model [which {\em a priori} would seem to be
unrelated to the present treatment of the ungauged action (\ref{eq:Zmass})], we
need only the first two orders of the expansion (\ref{eq:cum}). All higher
orders vanish. Thus, although we have not yet demonstrated a truly
non-perturbative boson-fermion equivalence for these mass terms, we feel
nevertheless that the present derivation represents a substantial improvement.

As for the restriction at intermediate steps to very definite distance
scales, we see here an effect which is bound to occur if the same
collective field technique is used to derive effective Lagrangians
(partly bosonized, or not) in higher dimensions. The analysis of
\cite{us} was in that sense rather particular in that for fermionic
couplings to just vector and axial vector currents in two dimensions, no
true field theoretic regularization was required beyond the calculation
of functional Jacobians. Here we have seen the more general machinery at
work. If we insist on a finite cut-off $\Lambda$, the fermionic theory
(\ref{eq:Zmass}) can only be partly be bosonized in a local fashion,
with rather unpleasant
non-local corrections of ${\cal{O}}(\Lambda^{-1})$. In addition, only in the
limit $g\to\infty$ do we immediately achieve a local action in the bosonic
representation.\footnote{Which, however, should be related through a
suitable non-local field redefinition to the form (29).}

It may be worthwhile to add a few comments on the computation of
chiral condensates, and in general fermionic correlation functions, of
the {\em original} fermionic theory (\ref{eq:Zmass}).
Even for finite $g$ the physical quantities should not
depend on this parameter. Thus, if one calculates expectation values (or Green
functions) in the effective $\theta$-theory, occurrences of $g$ should cancel
with contributions from the $\theta$-integration. Indeed, this is what happens.
Take as an example the chiral condensate $\langle \bar{\psi} P_\pm \psi
\rangle$ in a {\em free} theory, given by a functional derivative with respect
to $M_\pm$, which is subsequently set equal to zero. The
integration over $\theta$ has to be done with respect to the kinetic term
defined in ${\cal L}_{\theta}$, even for finite $g$. This yields
\beq
\langle \bar{\psi} P_\pm \psi \rangle = \frac{g e^\gamma}{\sqrt{\pi} 4\pi}
\langle e^{2i\theta } \rangle = \lim_{x\to 0}\frac{g e^\gamma}{\sqrt{\pi} 4\pi}
e^{- K_0(\mu |x|) + K_0 (\frac{g}{\sqrt{\pi}} |x|)} = \frac{\mu
e^\gamma}{4\pi} ,
\eeq
where $\mu$ is an infrared cut-off mass. This means that in going from the
formulation in terms of rotated fermions $\bar{\chi},\chi$ to the physical
fermion fields $\bar{\psi}$ and $\psi$, $g/\sqrt{\pi}$ is replaced by the
infrared
cut-off $\mu$,  which is finally taken to zero. One then recovers the correct
result for the chiral condensate of a free massless theory.

\subsection*{Acknowledgements}

We thank A. Wirzba for discussions.
H.B. Nielsen and, in parts, R. Sollacher acknowledge support by EEC grant
CS1-D430-C. The work of R. Sollacher has been supported mainly by the Deutsche
Forschungsgemeinschaft.

\bibliographystyle{unsrt}

\end{document}